\documentclass[pra,twocolumn,aps,showpacs,superscriptaddress,amsmath]{revtex4-1}

\usepackage{graphicx}
\usepackage{epstopdf}
\usepackage{bm}
\usepackage{dsfont}

\newcommand{\bra}[1]{\langle #1 |}
\newcommand{\ket}[1]{| #1 \rangle}
\newcommand{\braket}[2]{\langle #1 | #2 \rangle}

\newcommand{\us}{\uparrow}
\newcommand{\ds}{\downarrow}
\newcommand{\hlf}{\frac{1}{2}}
\newcommand{\veps}{\varepsilon}
\newcommand{\bs}[1]{\boldsymbol{#1}}

\begin{document}

\title{Engineering the dynamics of effective spin-chain models \\
for strongly interacting atomic gases}

\author{A.~G. Volosniev}
\affiliation{Department of Physics and Astronomy, Aarhus University, 
DK-8000 Aarhus C, Denmark} 

\author{D. Petrosyan}
\affiliation{Aarhus Institute of Advanced Studies, Aarhus University, 
DK-8000 Aarhus C, Denmark}
\affiliation{Institute of Electronic Structure and Laser, FORTH, 
GR-71110 Heraklion, Crete, Greece}

\author{M. Valiente}
\affiliation{SUPA, Institute of Photonics and Quantum Sciences, 
Heriot-Watt University, Edinburgh EH14 4AS, United Kingdom}

\author{D.~V. Fedorov}
\affiliation{Department of Physics and Astronomy, Aarhus University, 
DK-8000 Aarhus C, Denmark} 

\author{A.~S. Jensen}
\affiliation{Department of Physics and Astronomy, Aarhus University, 
DK-8000 Aarhus C, Denmark} 

\author{N.~T. Zinner}
\affiliation{Department of Physics and Astronomy, Aarhus University, 
DK-8000 Aarhus C, Denmark} 

\date{\today}

\begin{abstract}
We consider a one-dimensional gas of cold atoms with strong contact 
interactions and construct an effective spin-chain Hamiltonian for a 
two-component system. The resulting Heisenberg spin model can be 
engineered by manipulating the shape of the external confining potential 
of the atomic gas. We find that bosonic atoms offer more flexibility
for tuning independently the parameters of the spin Hamiltonian through 
interatomic (intra-species) interaction which is absent for fermions 
due to the Pauli exclusion principle. Our formalism can have important 
implications for control and manipulation of the dynamics of few- and many-body 
quantum systems; as an illustrative example relevant to quantum computation
and communication, we consider state transfer in the simplest non-trivial
system of four particles representing exchange-coupled qubits.  
\end{abstract}


\maketitle
    
\section{Introduction}

Interacting many-body quantum systems harbor many paradigmatic 
quantum phenomena, such as superconductivity and quantum magnetism, 
but are difficult to treat theoretically. For strong interparticle 
interactions, the usual perturbative and many numerical methods 
are inadequate, requiring more sophisticated approaches.
In the important case of one spatial dimension, relevant techniques 
include bosonization and the Tomonaga-Luttinger liquid theory 
\cite{haldane1981} and the numerically powerful density-matrix 
renormalization group methods \cite{white1992,scholl2005}. 

Cold atoms confined in magnetic and optical traps represent 
a remarkably clean and versatile system to simulate and study 
many-body physics under well-controlled conditions 
\cite{lewenstein2007,bloch2008,esslinger2010,Weitenberg2011,bloch2012}. 
Optical lattice potentials allow realization of the fundamental Hubbard 
model \cite{fisher1989,jaksch1998} in which quantum phase transition to 
the Mott insulator state with single atom per lattice site has been 
demonstrated \cite{greiner2002,jordans2008,Schneider2008}. 
The Mott-Hubbard insulator for a two-component system can be mapped 
onto the Heisenberg spin Hamiltonian \cite{kuklov2003,hild2014} 
facilitating studies of interacting spin models responsible for 
many key features of quantum magnetism \cite{auerbach1994}. 
One-dimensional (1D) systems of strongly interacting bosons 
\cite{paredes2004,kinoshita2004,haller2009,nicklas2014} and 
fermions \cite{zurn2012,wenz2013} have recently become experimentally 
accessible.

Experiments to simulate various lattice models with cold atoms typically
involve a weak trapping potential superimposed onto the optical 
lattice \cite{bloch2008}. The resulting potential deviates from 
an idealized homogeneous lattice, necessitating the use of the 
local-density approximation valid for a smooth trapping potential. 
Here we study an ensemble of cold alkali atoms in an external 
trapping potential having an arbitrary shape (not necessary lattice)
in the longitudinal direction but tightly confining the atoms in the 
transverse direction, realizing thereby an effective 1D system. 
A pair of internal atomic states from the ground-state hyperfine 
(Zeeman) manifold play the role of the spin-up and spin-down states. 
We show that such a 1D ensemble of strongly-interacting atoms with 
\textit{any} external confinement can, quite generally, be represented 
as a spin-chain. The strong contact interatomic interaction results 
in spatial localization of individual atoms within segments along 
the 1D trap, while the small but finite overlap between the wavefunctions
of neighboring atoms leads to an effective spin-exchange interaction.  
We construct an effective spin-$\hlf$ $XXZ$ model for a two-component
system and show that the parameters of the corresponding Hamiltonian 
sensitively depend on the shape of the confinement potential,
quantum statistics of the constituent atoms (bosons or fermions),
and the interatomic interaction (for bosons only).
We note an early relevant publication \cite{ogata1990} deriving an 
effective Heisenberg spin-$\hlf$ Hamiltonian for the homogeneous, 
large-$U$ Hubbard model at low-filling, and the very recent mapping 
of a multicomponent cold atomic gas in a harmonic trap onto a spin-chain
model \cite{deu2014}.

Our results open several possibilities for engineering stationary and 
dynamic quantum states of few- and many-body systems. As a revealing 
example amenable to analytic treatment, we consider the problem of 
quantum state transfer \cite{bose2003,StrRevs2007,nikolopoulos2014} 
in the simplest yet non-trivial case of four particles. We show that 
by optimal choice of the trapping potential and intra-species 
interactions between bosonic atoms, perfect transfer 
\cite{nikolopoulos2004,Christandl2004,kay2010} of a state 
of quantum bit -- qubit -- between the two ends of the spin chain 
can be attained. By contrast, fermions cannot accommodate perfect 
state transfer, unless they are subject to local (effective) magnetic 
fields.  

The paper is organized as follows. In the next Section, we 
demonstrate the equivalence of the eigenspectra of a two-component
ensemble on $N$ atoms in a 1D trap and a corresponding spin chain.
In Sec.~\ref{sec:contspchHam} we study the dependence of the
parameters of the effective spin Hamiltonian on the shape of
the external trapping potential for the atoms. Quantum dynamics 
and state transfer in engineered chains of four spins
is illustrated in Sec.~\ref{sec:engspch}, followed by
concluding remarks in Sec.~\ref{sec:conclud}.
In the Appendix we present perturbative derivation of the energy 
eigenvalues of the system using the approach of Ref.~\cite{volosniev2013},
analytic expressions for eigenspectrum of a four-spin system and dynamics 
of quantum state transfer, and derivation of an effective Heisenberg 
spin model in a magnetic field.

\section{Effective spin-chain model for $N$ atoms}
\label{sec:effspch}

Consider a 1D system of $N_{\us}$ particles of one kind (spin-up) and 
$N_{\ds}$ particles of another kind (spin-down) confined by an external 
trapping potential $V(x)$ with a characteristic length scale $L$. 
The total Hamiltonian for $N = N_{\us} + N_{\ds}$ particles is given by
\begin{eqnarray}
H &=& \sum_{\sigma=\us,\ds} \sum_{i=1}^{N_\sigma} \left[ h(x_{\sigma,i}) + 
\frac{g_{\sigma \sigma}\hbar^2}{mL}\sum_{i'>i}^{N_{\sigma}}\delta(x_{\sigma,i}-x_{\sigma,i'})
\right] \nonumber \\ & &
+ \frac{g_{\us \ds} \hbar^2}{mL}\sum_{i=1}^{N_{\us}}\sum_{i'=1}^{N_{\ds}}
\delta(x_{\us,i} - x_{\ds,i'}) , \label{Eq:Ham}
\end{eqnarray}
where 
\begin{equation}
h(x)= -\frac{\hbar^2}{2m}\frac{\partial^2} 
{\partial x^2}+\frac{\hbar^2}{mL^2}V(x/L)  \label{Eq:spHam}
\end{equation}
is the single particle Hamiltonian, 
$m$ is the mass assumed equal for all particles, and $x_{\us(\ds),i}$ 
denotes the position of the $i$th spin-up (spin-down) particle.
Throughout this paper, we use $L$ and $\veps \equiv \frac{\hbar^2}{mL^2}$
as units of length and energy, respectively. The zero-range interactions 
are parametrized by the dimensionless strengths $g_{\us \ds} \equiv g > 0$ 
and $g_{\ds \ds} = g_{\us \us} \equiv \kappa g$ with $\kappa>0$.
Hamiltonian~(\ref{Eq:Ham}) applies to both bosons and fermions, 
but the total wave function should be symmetric for bosons and 
antisymmetric for fermions. As a consequence, identical
(same-spin) fermions do not interact. 

We assume strong interactions, $g \gg 1$, and inspect the $N$-particle 
wavefunctions $\Phi(\{x_{\us,i},x_{\ds,i'}\})$ for various configurations 
$\{x_{\us,i},x_{\ds,i'}\}$ of atomic positions. There are in fact 
$\binom{N}{N_{\us}} = \frac{N!}{N_{\us}!N_{\ds}!}$ distinguishable
configurations with different ordering of atoms (spins), e.g., 
$x_{\us,1} < x_{\us,2} < x_{\ds,1} < \ldots < x_{\us,N_{\us}} < \ldots < x_{\ds,N_{\ds}}$.
In the limit of $1/g \to 0$, the requirement of finite energy implies 
that $\Phi$ should vanish whenever the coordinates of any two particles 
coincide, $x_{\us (\ds),i} = x_{\us (\ds),i'}$. This requirement can only be 
satisfied if $\Phi$ is proportional to the Slater determinant wavefunction 
for $N$ particles \cite{girardeau, volosniev2013}, which is a completely 
antisymmetric superposition of the products of different single-particle 
wavefunctions representing solutions of the single particle Hamiltonian $h(x)$. 
In what follows, we assume that the potential $V$ supports at least $N$ 
bound single particle levels, which are non-degenerate; the case of (partially)
degenerate spectrum can be treated similarly \cite{volosniev2013}.

Consider the Slater determinant wavefunction $\Phi_0$ composed of the $N$ 
lowest-energy single-particle eigenfunctions of $h(x)$. For $1/g \to 0$,
all $M(N_{\us},N_{\ds}) \equiv \binom{N_{\us} + N_{\ds}}{N_{\us}}$ configurations 
of atomic coordinates yield the same energy $E_0$ for 
$\Phi_0(\{x_{\us,i},x_{\ds,i'}\})$. We can expand the general 
$N$-particle eigenfunction as 
\begin{equation}
\Psi = \sum_{k=1}^{M(N_{\us},N_{\ds})} a_k \, \Pi_{k} 
\Phi_0(\{ : x_{\us,i} , x_{\ds,i'} : \} ) ,
\label{Eq:HamEigenstate}
\end{equation}
where $\{ : \! x_{\us,i} , x_{\ds,i'} \! : \} \equiv x_{\us,1} \! < \! \ldots \!
< x_{\us,N_{\us}} \! < x_{\ds,1} \! < \! \ldots \! < x_{\ds,N_{\ds}} $ and the sum 
is over all the permutations $\Pi_{k}$ of coordinates. Note that for $1/g=0$
any set of coefficients $a_k$ defines a legitimate ground-state, and we 
have in fact $M(N_{\us},N_{\ds})$ mutually-independent ground states of 
the same energy $E_0$. For small but finite $1/g$, the degeneracy 
of this ground state manifold is lifted, which follows from the 
Hellmann-Feynman theorem \cite{volosniev2013,volosniev2014}
\begin{eqnarray}
\frac{\partial E}{\partial g} &=& \kappa \sum_{\sigma=\us,\ds} 
\sum_{i=1}^{N_\sigma}\sum_{i'>i}^{N_\sigma} 
\langle \Psi|\delta(x_{\sigma,i}-x_{\sigma,i'})|\Psi\rangle 
\nonumber \\ & & 
+ \sum_{i=1}^{N_{\us}}\sum_{i'=1}^{N_{\ds}}
\langle \Psi|\delta(x_{\us,i}-x_{\ds,i'})|\Psi\rangle .
\label{Eq:HFTheorem}
\end{eqnarray}
Using the wavefunction of Eq.~(\ref{Eq:HamEigenstate}), we then obtain 
(see Appendix \ref{sec:app0}) the corresponding energy, to linear order in 
$1/g$ , as 
\begin{equation}
E = E_0 - \frac{\sum_{j=1}^{N-1} \frac{\alpha_j}{g} 
\left( A_{j} + \frac{2}{\kappa} C_{j} + \frac{2}{\kappa} D_{j}\right)}
{\sum_{k=1}^{M(N_{\downarrow},N_{\uparrow})} a_k^2},
\label{Eq:HamEigenEnergy}
\end{equation}
where  
\begin{eqnarray*}
A_{j} &=& \sum_{k=1}^{M(N_{\ds}-1,N_{\us}-1)} (a_{j|k}-b_{j|k})^2 , \\ 
C_{j} &=& \sum_{k=1}^{M(N_{\ds},N_{\us}-2)} c^2_{j|k} , \quad 
D_{j} = \sum_{k=1}^{M(N_{\ds}-2,N_{\us})} d^2_{j|k} ,
\end{eqnarray*}
for bosons, while $C_{j}=D_{j}=0$ for fermions. Here $a_{j|k}$ denote the $a_k$ 
coefficients in the expansion~(\ref{Eq:HamEigenstate}) multiplying terms 
$\Phi_0(\ldots <\underset{j}{x_{\us}}<\underset{j+1}{x_{\ds}}< \ldots)$ 
with $x_{\us}$ at position $j$ followed by $x_{\ds}$ at position $j+1$, 
while $b_{j|k}$ are the coefficients of
$\Phi_0(\ldots <\underset{j}{x_{\ds}}<\underset{j+1}{x_{\us}}< \ldots)$ 
with $x_{\us}$ and $x_{\ds}$ swapped.
Similarly, for identical bosons, $c_{j|k}$ denote the expansion 
coefficients in Eq.~(\ref{Eq:HamEigenstate}) in front of 
$\Phi_0(\ldots <\underset{j}{x_{\us}}<\underset{j+1}{x_{\us}}< \ldots)$,
while $d_{j|k}$ are the coefficients of
$\Phi_0(\ldots <\underset{j}{x_{\ds}}<\underset{j+1}{x_{\ds}}< \ldots)$.
Finally, the geometric factors $\alpha_j$ are solely determined 
by the confining potential through $\Phi_0$ as
\begin{equation}
\alpha_j = 
\frac{\int \prod_{i=1}^{N_{\uparrow}} d x_{\us,i} \prod_{i'=1}^{N_{\ds}}d x_{\ds,i'} 
\left| \dfrac{\partial \Phi_0}{\partial x_{\ds,1}} \right|^2 
\delta(x_{\us,j} - x_{\ds,1})} 
{\int \prod_{i=1}^{N_{\uparrow}} d x_{\us,i} \prod_{i'=1}^{N_{\ds}}d x_{\ds,i'}  
|\Phi_0(\{ : \! x_{\us,i} , x_{\ds,i'} \! : \})|^2 }, \label{Eq:alphas}
\end{equation}
where in $\Phi_0$ in the numerator the spin-down atom $x_{\ds,1}$ is placed 
at position $j+1$ following $j$ spin-up atoms $x_{\us,1} \ldots x_{\us,j}$.
Below we deal mostly with bosons as they can also reproduce fermions
in the limit of $\kappa \to \infty$.

We now demonstrate that Hamiltonian (\ref{Eq:Ham}) for $N = N_{\us} + N_{\ds}$
strongly-interacting particles, $1/g \ll 1$, can be mapped onto the 
spin-$\hlf$ $XXZ$ Hamiltonian of the form 
\begin{equation}
H_s = E_0 \mathbf{I} - \hlf \sum_{j=1}^{N-1} 
\bigg[J_j (\bs{\sigma}^j \bs{\sigma}^{j+1} - \mathbf{I} ) -
\frac{2 J_j}{\kappa} (\sigma_z^j \sigma_z^{j+1}+\mathbf{I} ) \bigg],
\label{Eq:HamSpin}
\end{equation}
where $\mathbf{I}$ is the identity matrix, 
$\bs{\sigma}^j=(\sigma^j_x,\sigma^j_y,\sigma^j_z)$ are the Pauli 
matrices acting on the spin at site $j$, 
and $J_j$ are position-dependent interaction coefficients. 
Note that $H_s$ conserves the total spin projection, $\Sigma_z = N_{\us}-N_{\ds}$.

Any eigenstate of \eqref{Eq:HamSpin} can be expanded in terms of the 
spin permutations $\Pi_{k}$ as
\begin{equation}
\ket{\Psi} = \sum_{k=1}^{M(N_{\us},N_{\ds})} a_{k} \, \Pi_{k} 
\ket{\us_1 \ldots\us_{N_{\us}} \, \ds_1 \ldots\ds_{N_{\ds}}} .
\end{equation}
Consider the energy expectation value $\bra{\Psi} H_s \ket{\Psi}$.
Using the swap operator 
$\mathbf{P}_{j,j+1} = \hlf (\bs{\sigma}^j \bs{\sigma}^{j+1} + \mathbf{I})$,
we find that the non-zero contributions to 
$\hlf \bra{\Psi} \bs{\sigma}^j \bs{\sigma}^{j+1} - \mathbf{I} \ket{\Psi}$ 
are 
\begin{eqnarray*}
& [a_{j|k} \bra{\Phi_{j|k}} + b_{j|k} \bra{\Phi_{j|k}} \mathbf{P}_{j,j+1}] 
(\mathbf{P}_{j,j+1} - \mathbf{I}) \\ 
& \times [a_{j|k} \ket{\Phi_{j|k}} + b_{j|k} \mathbf{P}_{j,j+1} \ket{\Phi_{j|k}}] \\
& = -a_{j|k}^2 - b_{j|k}^2 + 2 a_{j|k} b_{j|k} ,  
\end{eqnarray*}
where  $\ket{\Phi_{j|k}} 
\equiv \ket{\ldots \underset{j}\us \underset{j+1}\ds \ldots}$.
Assuming normalization $\braket{\Psi}{\Psi} = 1$, we then obtain
\begin{equation}
\bra{\Psi} H_s \ket{\Psi} = E_0 + \sum_{j=1}^{N-1} J_j 
\left(A_{j} + \frac{2}{\kappa} C_{j} + \frac{2}{\kappa} D_{j} \right), 
\label{Eq:HamSpinExpeVal}
\end{equation}
with $A_{j}$, $C_{j}$ and $D_{j}$ having the same meaning as above. 
Comparison of Eqs.~(\ref{Eq:HamEigenEnergy}) and (\ref{Eq:HamSpinExpeVal}) 
reveals that, to linear order in $1/g$, the eigenvalue problem for 
the Hamiltonians (\ref{Eq:Ham}) and (\ref{Eq:HamSpin}) is the same, 
with the corresponding spin-spin interaction coefficients given by 
$J_j = -\alpha_j/g$. Note that since the overlap integrals $\alpha_j$ 
are always positive, the coefficients $J_j$ are negative, which is
to be expected for strongly repulsive interatomic interaction $g >0$.
(This is similar to the optical lattice setup \cite{kuklov2003},
where the spin-spin interactions are mediated by virtual intermediate
two-atom states having higher energy and therefore pushing the energies
of single-atom states down). For fermions or hard-core bosons, 
$\kappa \to \infty$, Eq.~(\ref{Eq:HamSpin}) becomes the $XXX$ Hamiltonian, 
while in the special case of bosons with $\kappa = 2$, it reduces to 
the $XX$ model Hamiltonian. These results are summarized in Table~\ref{table}.
In Ref.~\cite{deu2014} Deuretzbacher {\it et al.} also arrive at a 
spin model Hamiltonian for a harmonically trapped two-component atomic 
gas of fermions with $\kappa\to\infty$ and bosons with $\kappa=1$.

\begin{table}[t]
 \caption{Effective Heisenberg spin models 
for strongly interacting atoms in 1D traps.  
\label{table}}
 \begin{ruledtabular}
  \begin{tabular}{ccccccc}
   & Spin-$\hlf$ model & Constituents & $\kappa$  & \\[2pt]\hline
   & $XXZ$ & bosons & $0 < \kappa < \infty$  \\
   & $XXX$ & bosons or fermions & $\kappa \to \infty$  \\
   & $XX$  & bosons & $\kappa=2$ 
  \end{tabular}
 \end{ruledtabular}
\end{table}

For concreteness, we have contemplated so far only the ground state 
energy manifold of Hamiltonian~(\ref{Eq:Ham}). Yet, precisely the same 
arguments apply to any $n$th excited state manifold which can be 
represented by a corresponding $XXZ$ Hamiltonian (\ref{Eq:HamSpin}) 
disconnected from all the other energy manifolds, each located in 
the vicinity of energy $E_n$ of the corresponding Slater determinant 
wavefunction $\Phi_n$. This of course holds for small enough timescales
when we can neglect energy relaxations, finite temperature and other 
effects causing transitions between different energy manifolds $E_n$
of the system. 

\section{Controlling the spin-chain Hamiltonian}
\label{sec:contspchHam}

\begin{figure}[b]
\centerline{\includegraphics[width=6.0cm]{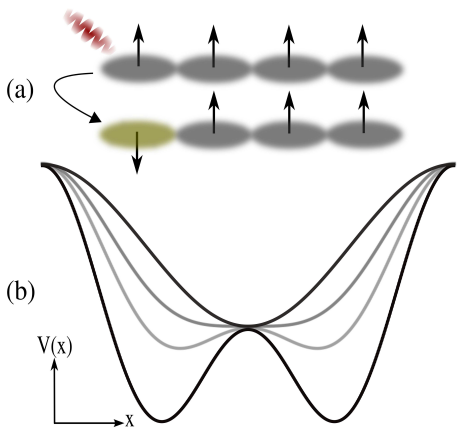}}
\caption{(Color online)
(a) A system of four atoms in a 1D trap 
is initialized by changing the internal state (flipping spin) 
of one of the atoms. 
(b) Trapping potential of Eq.~\eqref{eq:pot} for $V_0 = 50\veps$ 
and $u=(0,1,2,4) \times u_{\mathrm{p}}$ (top to bottom) with 
$u_{\mathrm{p}} \simeq 12.5 \veps$.}
\label{fig:potential}
\end{figure}

The above analysis attests to the possibility of tuning the interspin 
couplings $J_j$ and anisotropy of the effective Hamiltonian $H_s$, 
Eq.~(\ref{Eq:HamSpin}), through the trapping potential $V(x)$ and 
interparticle interactions $g \gg 1$ and $\kappa >0$. As an illustration,
consider a relatively simple yet non-trivial system of four particles 
confined in a symmetric double-well trap of the form 
(see Fig.~\ref{fig:potential})
\begin{equation}
V(x)= - V_0 \sin^2\left[\hlf (x+1) \pi\right]
- u \sin^2\left[(x+1)\pi\right], \label{eq:pot}
\end{equation}
Varying parameter $u \geq 0$, we may change the potential 
whose depth $V_0=50\veps$ is chosen large enough to accommodate 
at least four well-localized single particle levels. 
This allows us to restrict the problem to $x \in [-1,1]$ with hard 
wall boundaries at $|x|=1$ and obtain accurate wavefunctions $\Phi_0$. 

\begin{figure}[t]
\centerline{\includegraphics[width=8.7cm]{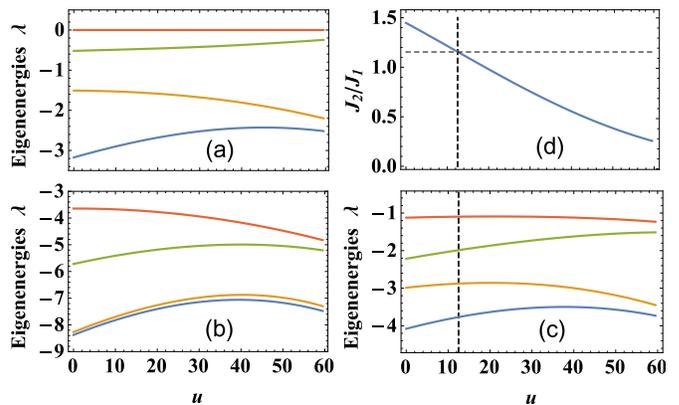}}
\caption{(Color online)
Energy eigenvalues $\lambda_n$ of $H_s$ (less the $E_0 \mathbf{I}$ term)
vs $u$ of Eq.~(\ref{eq:pot}), for $N_{\ds}=1$, $N_{\us}=3$ and 
(a) $\kappa \to \infty$ (fermions), (b) $\kappa = \hlf$, and (c) $\kappa = 2$. 
The ratio $J_2/J_1$ of the coupling constants is shown in (d), 
with dashed lines marking $J_2/J_1=\sqrt{4/3}$ and $u=u_\textrm{p}$,
corresponding to equidistant spectrum in (c). 
$\lambda$'s and $u$ are in units of $\veps$ and $g=100$. }
\label{fig:HamSpinEigVals}
\end{figure}

We assume that three of the particles are prepared in the internal 
(spin) state $\ket{\!\us}$ and the fourth is in state $\ket{\!\ds}$
[Fig.~\ref{fig:potential}(a)]. The system is then non-trivial since 
the interspin coupling coefficients $J_1$ ($=J_3$) and $J_2$ can be tuned 
independently, which is not possible for less than four particles in 
a symmetric trap. In Fig.~\ref{fig:HamSpinEigVals} we show the 
dependence of energy eigenvalues $\lambda_n$ of $H_s$ and the ratio 
$J_2/J_1$ on the parameter $u$ of the potential of Eq.~(\ref{eq:pot}). 
Clearly, for $u \ll V_0$ the potential $V(x)$ is nearly harmonic, 
leading to larger overlap of the wavefunctions of the particles
in the middle of the trap, which results in $J_2/J_1 \simeq 1.4$. 
Increasing $u$ we decrease the overlap and thereby the coupling 
strength $J_{2}$ relative to $J_{1,3}$, see Fig.~\ref{fig:HamSpinEigVals}(d).
For very large $u \gg V_0$, the system splits into two non-interacting 
parts with vanishing coupling $J_2$ in the middle and doubly 
degenerate eigenvalues. This tendency can be seen in 
Fig.~\ref{fig:HamSpinEigVals}(a)-(c), where we use three representative
values of $\kappa$. The fermionic case of $\kappa \to \infty$ 
corresponds to isotropic spin Hamiltonian (see Appendix~\ref{subsec:appAFerm}). 
In the bosonic case with $\kappa < 1$, the interaction with 
the impurity (spin-down) particle is stronger than the interaction 
between identical (spin-up) particles. As a result, the pair 
of lowest energy eigenstates, corresponding approximately to 
configurations $\ket{\!\ds \us \us \us} \pm \ket{\!\us \us \us \ds}$
with the impurity particle at the boundary, are almost completely
decoupled from the other configurations, and therefore are nearly 
degenerate (see Appendix~\ref{subsec:appABos}), which was also 
discussed in \cite{zinner2013}.    
The case of $\kappa = 2$ corresponding to the $XX$ model is of special 
interest in the following. As seen in Fig.~\ref{fig:HamSpinEigVals}(d), 
by choosing $u=u_{\mathrm{p}} \simeq 12.5 \veps$ we obtain for the ratio 
of the coupling strengths $J_2/J_1 = \sqrt{4/3}$ leading to the equidistant
eigenspectrum in Fig.~\ref{fig:HamSpinEigVals}(c).

\section{Quantum dynamics in engineered spin-chains}
\label{sec:engspch}

The possibility to realize various spin chain Hamiltonians with cold 
trapped atoms can have important implications for quantum simulations 
and computation \cite{lewenstein2007,bloch2012}. A potentially useful 
application of quantum dynamics in engineered spin chains can be state 
transfer in small quantum networks \cite{bose2003,StrRevs2007,nikolopoulos2014}.
Faithful transfer of quantum states is a prerequisite for achieving scalable 
quantum information processing in lattice-based schemes where qubit-qubit 
interactions are typically short range and implementing quantum logic gates 
between distant qubits requires interconnecting them via quantum channels 
represented by tunable spin chains \cite{petrosyan2010}.

In its standard form \cite{bose2003,StrRevs2007,petrosyan2010},
the quantum state transfer protocol involves preparing the spin chain 
in a dynamically passive state, e.g., $\ket{\!\us \us \ldots \us \us}$,
and then initializing at time $t_\mathrm{in} = 0$ the first spin with 
the qubit state $\ket{\psi} = \alpha \ket{\!\us} + \beta \ket{\!\ds}$ 
to be transferred. Ideal transfer would imply that at a well-defined 
time $t_\mathrm{out}$ the last spin of the chain is in state $\ket{\psi}$ 
(up to a certain relative phase $\phi_0$ between the amplitudes 
of $\ket{\!\us}$ and $\ket{\!\ds}$).
Since for the qubit state $\ket{\!\us}$ the spin chain remains in the 
passive state, our aim is to maximize the probability of attaining state 
$\ket{\!\us \us \ldots \us \ds}$ at time $t_\mathrm{out}$ given that at 
time $t=0$ its state was  $\ket{\Psi(0)} = \ket{\!\ds \us \ldots \us \us}$. 
We thus define the fidelity of state transfer as 
\begin{equation}
F(t) \equiv |\braket{\Psi(t)}{\!\us \us \ldots \us \ds}|^2 .
\end{equation}

The chain of four spins initialized as shown in Fig.~\ref{fig:potential}(a) 
represents the smallest non-trivial system in which achieving perfect 
state transfer, $F(t_\mathrm{out}) = 1$, requires judicious choice of 
the parameters of Hamiltonian $H_s$. Indeed, in a two-state system, 
resonant coupling $J_1$ between $\ket{\!\ds \us}$ and $\ket{\!\us \ds}$
amounts to complete Rabi oscillations, while in a three-state system 
with degenerate initial $\ket{\!\ds \us \us}$ and final 
$\ket{\!\us \us \ds}$ states, and not too large energy offset of 
the intermediate state $\ket{\!\us \ds \us}$, any $J_1$ and $J_2$
result in effective Rabi oscillations between the initial and 
final states \cite{nikolopoulos2004,Christandl2004}.
The necessary and sufficient condition for perfect state transfer 
in a spin chain of any length $N$ is a commensurate spectrum 
of $H_s$ \cite{kay2010}, namely 
$e^{-i \lambda_n t_{\mathrm{out}}} = (-1)^n e^{i \phi}$ with some $\phi$,
the equidistant spectrum, $\lambda_{n+1} - \lambda_n = \Delta \lambda \, 
\forall \, n$, being optimal \cite{yung2006} in terms of the fastest 
transfer time $t_{\mathrm{out}} = \hbar \pi /\Delta \lambda$. 
In the case of the $XX$ Hamiltonian, perfect and optimal state 
transfer is realized by choosing the coupling constants as 
$J_j=J_0 \sqrt{(N-j)j}$ \cite{Christandl2004,nikolopoulos2004}
resulting in $t_{\mathrm{out}} = \hbar \pi /2J_0$. For our system
of $N=4$ spins, this corresponds to $J_2/J_1 = \sqrt{4/3}$
[see Fig.~\ref{fig:HamSpinEigVals}(c),(d)] and transfer time 
$t_{\mathrm{out}} = \hbar \pi /J_2$.

\begin{figure}
\centerline{\includegraphics[width=7.5cm]{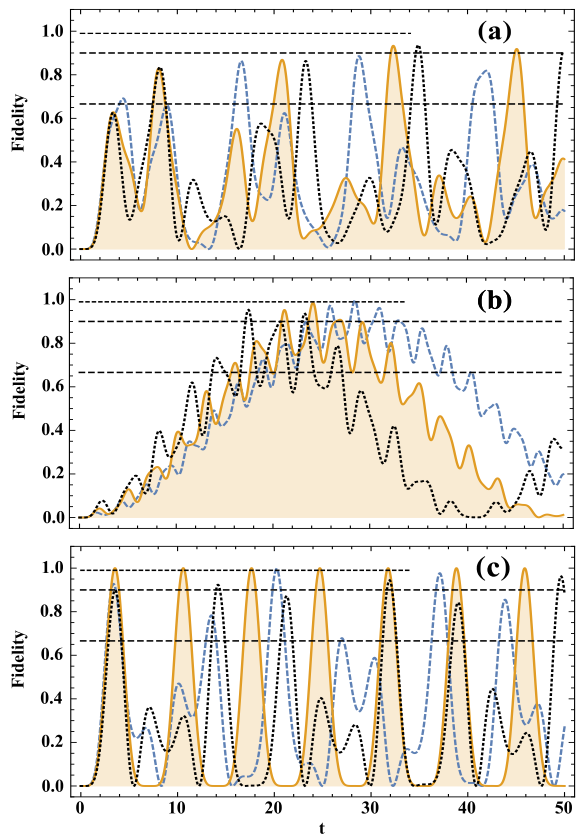}}
\caption{(Color online)
Fidelity $F(t)$ of state transfer in a four-spin system with 
(a) $\kappa \to \infty$ (fermions), (b) $\kappa = \hlf$, and (c) $\kappa = 2$ 
[cf. Fig.~\ref{fig:HamSpinEigVals}(a), (b), and (c)], 
for $u=0$ (dashed, blue), $u_\textrm{p}$ (solid, orange), 
and $2 u_\textrm{p}$ (dotted, black). 
For visual aid, the values of $F=2/3,0.9,0.99$ are marked with thin 
dashed horizontal lines. Time is in units of $\hbar/\veps$.}
\label{fig:fidelity}
\end{figure}

In Fig. \ref{fig:fidelity} we show the time-dependence of fidelities 
$F(t)$ of state transfer for the same values of $\kappa$ as in 
Fig.~\ref{fig:HamSpinEigVals}. Due to incommensurate spectrum, 
the fermionic ($XXX$) case $\kappa \to \infty$ without external
magnetic field (see below) cannot realize perfect state transfer 
for any $u$. This we prove in Appendix~\ref{sec:appA}, where we also 
show that bosons with $\kappa = 1$ yield the same fidelity as fermions
in Fig.~\ref{fig:fidelity}(a). In the bosonic case with $\kappa < 1$, 
we observe in Fig.~\ref{fig:fidelity}(b) a slow (third order in $J_j$)
transition between the degenerate initial $\ket{\!\ds \us \us \us}$ and 
final $\ket{\!\us \us \us \ds}$ states via nonresonant intermediate 
states $\ket{\!\us \ds \us \us}$ and $\ket{\!\us \us \ds \us}$ 
(see Appendix~\ref{subsec:appABos} for details).
Finally, the perfect, optimal state transfer is realized in the
$\kappa = 2$ ($XX$) case with $u=u_\textrm{p}$, Fig.~\ref{fig:fidelity}(c), 
as expected,

We note finally that the $XXX$ Hamiltonian can in principle be modified
by a spatially inhomogeneous (effective) magnetic field $B(x)\sigma_z$ 
resulting in 
\begin{equation}
\tilde{H}_s = H_s + \sum_{j=1}^{N} h_j \sigma^j_z ,
\end{equation}
as shown in Appendix~\ref{sec:appB}. 
Then, for $N_{\ds} =1$, an appropriate choice of the local fields, 
$h_{1,N} = J_{1,N-1}$ and $h_{j = 2,\ldots, N-1} = J_{j-1} + J_j$, will 
equalize the diagonal elements of $\tilde{H}_s$, turning it into the 
$XX$ Hamiltonian, which, with proper interspin coupling coefficients $J_j$
determined by the trapping potential $V(x)$, can realize perfect state transfer.

\section{Conclusions}
\label{sec:conclud}

We have shown that a two component system of strongly-interacting 
atoms in a 1D trap can be represented as a spin chain described 
by the $XXZ$ model Hamiltonian. Quite generally, any number of atoms $N$
in an arbitrary trapping potential -- not necessary spatially periodic --
is amenable to such a representation. To obtain the corresponding
spin-chain Hamiltonian (\ref{Eq:HamSpin}), one has to construct the Slater 
determinant wavefunction $\Phi_0$ from $N$ single-particle eigenfunctions 
in the trap of a given form $V(x)$ and then calculate the overlap integrals 
$\alpha_j$ of Eq.~(\ref{Eq:alphas}) yielding the spin-spin interaction 
coefficients $J_j = -\alpha_j/g$. In turn, the shape of the trapping 
potential determines the parameters of the resulting Hamiltonian,
which permits (reverse) engineering of the desired many-body states 
and dynamics of the effective spin chain. 

Our formalism, while applicable to particles with strong contact 
interactions, scales favorably with the particle number $N$. 
Moreover, our approach is easily extendable to multicomponent 
(spin $s > 1/2$) systems analogous to spin chain models with $SU(2s+1)$ 
symmetry. Chains of coupled qudits of dimension $2s+1>2$ exhibit higher 
quality of entanglement transfer \cite{bayat2007}.

The experimental context of our study is cold alkali atoms, e.g., Rb or Li,
in small traps of dimension $L \sim 1\:\mu$m realized by far-detuned focused 
laser beams or optical lattices. The corresponding energy scale is then
$\veps/\hbar \sim 1-10\:$kHz, while strong interactions $g \gg 1$ occur
near Feshbach resonances in external magnetic fields.
Tailoring magnetic fields on the scale of $L$ could be difficult. Instead, 
appropriately detuned, tightly focused laser beams can mimic spatially 
inhomogeneous magnetic fields through differential Stark shifts of the 
hyperfine (Zeeman) atomic levels, and can induce Raman transitions between
(spin) states of individual atoms to prepare, initialize and read-out the 
state of the system as required \cite{Weitenberg2011,bloch2012}.
 
\begin{acknowledgments}
A.G.V. and N.T.Z. thank the Institute for Nuclear Theory in Seattle 
for hospitality during the program INT-14-1 'Universality in Few-Body Systems'. 
Useful discussions with Jacob Sherson and G.M. Nikolopoulos are gratefully
acknowledged. This work was funded by the Danish Council for Independent 
Research DFF Natural Sciences and the DFF Sapere Aude program.
\end{acknowledgments}

\appendix

\section{Perturbative derivation of the energy eigenvalues, 
Eq.~(\ref{Eq:HamEigenEnergy})}
\label{sec:app0}

Here we outline the derivation of the energy eigenvalues of Hamiltonian 
(\ref{Eq:Ham}). The corresponding eigenvalue problem is defined by 
the Schr{\"o}dinger equation 
\begin{equation}
\sum_{\sigma=\us,\ds}\sum_{i=1}^{N_{\sigma}} h(x_{\sigma,i}) \Psi=E\Psi ,
\label{EqA:EigProp}
\end{equation} 
supplemented with the boundary conditions at the contact
positions of any two particles, 
\begin{equation}
\left(\frac{\partial \Psi}{\partial x_{\sigma,i}}-
\frac{\partial \Psi}{\partial x_{\sigma',i'}}\right)\bigg|^{x_{\sigma,i}-
x_{\sigma',i'}=0^+}_{x_{\sigma,i}-x_{\sigma',i'}=0^-}
=2g_{\sigma \sigma' }\Psi(x_{\sigma,i}=x_{\sigma',i'}).
\label{EqA:BoundConds}
\end{equation}
The dependence of the energy $E$ on the interaction strength $g$ can be 
inferred from the Hellmann-Feynman theorem \cite{volosniev2013,volosniev2014},
Eq.~(\ref{Eq:HFTheorem}),
\begin{eqnarray}
\frac{\partial E}{\partial g} &=& \kappa \sum_{\sigma=\us,\ds} 
\sum_{i=1}^{N_\sigma}\sum_{i'>i}^{N_\sigma} 
\langle \Psi|\delta(x_{\sigma,i}-x_{\sigma,i'})|\Psi\rangle 
\nonumber \\ & & 
+ \sum_{i=1}^{N_{\us}}\sum_{i'=1}^{N_{\ds}}
\langle \Psi|\delta(x_{\us,i}-x_{\ds,i'})|\Psi\rangle .
\label{EqA:HFTheorem}
\end{eqnarray}
Combining Eqs. (\ref{EqA:BoundConds}) and (\ref{EqA:HFTheorem}), we obtain 
for small $1/g$
\begin{equation}
\frac{\partial E}{\partial g} = \frac{K_{\us\ds}}{g^2}+\frac{K_{\us\us}}{\kappa g^2}
 + \frac{K_{\ds\ds}}{\kappa g^2} + O(1/g^2), \label{EqA:dEdg}
\end{equation}
with the interaction parameters $K_{\sigma \sigma'}$ given by
\begin{widetext}
\begin{eqnarray}
K_{\us \ds} &=& \lim_{g\to\infty} 
\frac{\sum_{i=1}^{N_{\us}} \sum_{i'=1}^{N_{\ds}} 
\int \prod_{j=1}^{N_{\us}} d x_{\us,j} \prod_{j'=1}^{N_{\ds}} d x_{\ds,j'} 
\left| \left( \dfrac{\partial \Psi}{\partial x_{\ds,i}}
- \dfrac{\partial \Psi}{\partial x_{\us,i'}} \right) 
\bigg|^{x_{\ds,i}-x_{\us,i'}=0^+}_{x_{\ds,i}-x_{\us,i'}=0^-}\right|^2 \delta(x_{\ds,i}-x_{\us,i'})}
{4 \int \prod_{i=1}^{N_{\uparrow}} d x_{\us,i} \prod_{i'=1}^{N_{\ds}}d x_{\ds,i'}  |\Psi|^2 }
, \label{Eq:SlopeUD} \\
K_{\sigma \sigma} &=& \lim_{g\to\infty} 
\frac{\sum_{i=1}^{N_{\sigma}} \sum_{i' >i}^{N_{\sigma}} 
\int \prod_{j=1}^{N_{\us}} d x_{\us,j} \prod_{j'=1}^{N_{\ds}} d x_{\ds,j'} 
\left| \left( \dfrac{\partial \Psi}{\partial x_{\sigma,i}}
- \dfrac{\partial \Psi}{\partial x_{\sigma,i'}} \right) 
\bigg|^{x_{\sigma,i}-x_{\sigma,i'}=0^+}_{x_{\sigma,i}-x_{\sigma,i'}=0^-}\right|^2 
\delta(x_{\sigma,i}-x_{\sigma,i'})}
{4 \int \prod_{i=1}^{N_{\uparrow}} d x_{\us,i} \prod_{i'=1}^{N_{\ds}}d x_{\ds,i'}  |\Psi|^2 }
, \label{Eq:SlopeUUDD}
\end{eqnarray}
with $\sigma = \us$ or $\ds$. Apparently, different wavefunctions $\Psi$
with the corresponding combinations of $a_k$ in Eq.~(\ref{Eq:HamEigenstate})
lead to different values of $K_{\sigma \sigma'}$ which lifts the degeneracy 
of the spectrum. By integrating Eq.~(\ref{EqA:dEdg}) with respect to $g$, we 
obtain the perturbative expansion (\ref{Eq:HamEigenEnergy}) used in 
Sec.~\ref{sec:effspch}.

\section{Static and dynamic properties of the effective spin model 
with $N_{\ds}=1$ and $N_{\us}=3$}
\label{sec:appA}

Here we present analytic expressions for the eigenvalues and eigenvectors
of the spin Hamiltonian $H_s$ for four particles, three of which are in 
one internal state (spin-up) and the other one is in a different internal 
state (spin-down), and analyze the state transfer dynamics.

In the basis of $\{ \ket{\!\ds \us \us \us}, \ket{\!\us \ds \us \us}, 
\ket{\!\us \us \ds \us}, \ket{\!\us \us \us \ds} \}$, 
the Hamiltonian in Eq.~(\ref{Eq:HamSpin}) can be cast in the matrix form
\begin{equation}
H_s - E_0 {\bf I} =
 \begin{pmatrix}
  J_1+\frac{2J_1}{\kappa} + \frac{2J_2}{\kappa} & -J_1 & 0 & 0 \\
  -J_1 & J_1 + \frac{2J_1}{\kappa} + J_2 & -J_2 & 0 \\
  0  & -J_2  & J_1 + \frac{2J_1}{\kappa} + J_2 & -J_1  \\
  0 & 0 & -J_1 & J_1 + \frac{2J_1}{\kappa} + \frac{2J_2}{\kappa}
 \end{pmatrix} ,
\label{SupEq:HamSpinMartix}
\end{equation}
where the $E_0 {\bf I}$ term yields a trivial common energy shift for 
all spin configurations and can therefore be dropped. For finite $\kappa$, 
Eq.~(\ref{SupEq:HamSpinMartix}) describes bosons, and we see that 
for $\kappa=2$ all the diagonal elements of the matrix are the same,
which is in fact the Heisenberg $XX$ model.  The fermionic limit 
$\kappa \to \infty$ corresponds to the isotropic $XXX$ model. 

The eigenvalues of Eq.~(\ref{SupEq:HamSpinMartix}) are
\begin{eqnarray*}
\lambda_1 &=& \frac{2 J_1 + \kappa J_1 + J_2 - \sqrt{\kappa^2 J_1^2 + J_2^2}}{\kappa} , \\
\lambda_2 &=& \frac{2 J_1 + \kappa J_1 + J_2 + \sqrt{\kappa^2 J_1^2 + J_2^2}}{\kappa} , \\
\lambda_3 &=& \frac{2 J_1 + \kappa J_1 + J_2 + \kappa J_2 
- \sqrt{\kappa^2 J_1^2 + J_2^2 - 2 \kappa J_2^2 + \kappa^2 J_2^2}}{\kappa} , \\
\lambda_4 &=& \frac{2 J_1 + \kappa J_1 + J_2 + \kappa J_2 
+ \sqrt{\kappa^2 J_1^2 + J_2^2 - 2 \kappa J_2^2 + \kappa^2 J_2^2}}{\kappa} ,
\end{eqnarray*}
with the corresponding (non-normalized) eigenvectors
\begin{eqnarray*}
\ket{\Psi_1} &=& \{ 1, 
  \frac{J_2 + \sqrt{\kappa^2 J_1^2 + J_2^2}}{\kappa J_1}, 
  \frac{J_2 + \sqrt{\kappa^2 J_1^2 + J_2^2}}{\kappa J_1}, 
  1 \}, \\
\ket{\Psi_2} &=& \{ 1, 
  \frac{J_2 - \sqrt{\kappa^2 J_1^2 + J_2^2}}{\kappa J_1}, 
  \frac{J_2 - \sqrt{\kappa^2 J_1^2 + J_2^2}}{\kappa J_1}, 
  1 \}, \\
\ket{\Psi_3} &=& \{-1, 
  -\frac{J_2 - \kappa J_2 + \sqrt{\kappa^2 J_1^2 + J_2^2 - 2 \kappa J_2^2 + \kappa^2 J_2^2}}{\kappa J_1}, 
   \frac{J_2 - \kappa J_2 + \sqrt{\kappa^2 J_1^2 + J_2^2 - 2 \kappa J_2^2 + \kappa^2 J_2^2}}{\kappa J_1}, 
   1 \} ,  \\
\ket{\Psi_4} &=& \{-1, 
  -\frac{J_2 - \kappa J_2 - \sqrt{\kappa^2 J_1^2 + J_2^2 - 2 \kappa J_2^2 + \kappa^2 J_2^2}}{\kappa J_1}, 
   \frac{J_2 - \kappa J_2 - \sqrt{\kappa^2 J_1^2 + J_2^2 - 2 \kappa J_2^2 + \kappa^2 J_2^2}}{\kappa J_1}, 
  1\} .
\end{eqnarray*}

\subsection{Fermions} 
\label{subsec:appAFerm}

In the limit of $\kappa \to \infty$, the ordered eigenvalues 
and normalized eigenvectors reduce to
\begin{eqnarray*} 
\lambda^{(\mathrm{f})}_1 = 0, & \quad  & \ket{\Psi_1^{(\mathrm{f})}} 
= \frac{1}{2} \{1,1,1,1\} ; \\
\lambda^{(\mathrm{f})}_2 = J_1 + J_2 - \sqrt{J_1^2 + J_2^2}, & & 
\ket{\Psi_2^{(\mathrm{f})}} = \frac{1}{2 \sqrt{J_1^2+J_2^2 - J_2 \sqrt{J_1^2+J_2^2} }}\{-J_1,J_2-\sqrt{J_1^2+J_2^2}, -J_2+\sqrt{J_1^2+J_2^2}, J_1\} ; \\
\lambda^{(\mathrm{f})}_3 = 2J_1 , & & \ket{\Psi_3^{(\mathrm{f})} } = \frac{1}{2} \{1,-1,-1,1\} ; \\
\lambda^{(\mathrm{f})}_4 = J_1 + J_2 + \sqrt{J_1^2 + J_2^2},  &  &  
\ket{\Psi_4^{(\mathrm{f})}}= \frac{1}{2 \sqrt{J_1^2+J_2^2 + J_2 \sqrt{J_1^2+J_2^2} }}\{-J_1,J_2+\sqrt{J_1^2+J_2^2},-J_2-\sqrt{J_1^2+J_2^2},J_1\} .
\end{eqnarray*}

Our aim is to transfer the initial state 
$\ket{\Psi^{(\mathrm{f})}_{\mathrm{in}}} = \ket{\!\ds \us \us \us}$, 
which evolves in time as 
\begin{eqnarray*}
\ket{\Psi^{(\mathrm{f})}(t)} &=& \frac{1}{2} \ket{\Psi_1^{(\mathrm{f})}} e^{-i \lambda^{(\mathrm{f})}_1 t} 
- \frac{J_1}{2 \sqrt{J_1^2+J_2^2 - J_2 \sqrt{J_1^2+J_2^2} }} \ket{\Psi_2^{(\mathrm{f})}} e^{-i \lambda^{(\mathrm{f})}_2 t} \\ & &
+ \frac{1}{2} \ket{\Psi_3^{(\mathrm{f})}} e^{-i \lambda^{(\mathrm{f})}_3 t} 
- \frac{J_1}{2 \sqrt{J_1^2+J_2^2 + J_2 \sqrt{J_1^2+J_2^2} }} \ket{\Psi_4^{(\mathrm{f})}} e^{-i \lambda^{(\mathrm{f})}_4 t},
\end{eqnarray*}
to the final state 
\begin{eqnarray*}
\ket{\Psi^{(\mathrm{f})}_{\mathrm{out}}} = \ket{\!\us \us \us \ds} &=& 
\frac{1}{2} \ket{\Psi_1^{(\mathrm{f})}}
+ \frac{J_1}{2 \sqrt{J_1^2+J_2^2 - J_2 \sqrt{J_1^2+J_2^2} }} \ket{\Psi_2^{(\mathrm{f})}} 
+ \frac{1}{2} \ket{\Psi_3^{(\mathrm{f})}}
+ \frac{J_1}{2 \sqrt{J_1^2+J_2^2 + J_2 \sqrt{J_1^2+J_2^2} }} \ket{\Psi_4^{(\mathrm{f})}} .
\end{eqnarray*}
\end{widetext}
The necessary and sufficient conditions for this are 
\begin{equation}
(\lambda^{(\mathrm{f})}_{n+1}-\lambda^{(\mathrm{f})}_{n}) t_{\mathrm{out}} = (2m_n+1) \pi,
\end{equation}
where $m_n$ are some positive integers and $t_{\mathrm{out}}$ is a transfer time. 
This leads to two equations for $m_{1,2,3}$
\begin{eqnarray*}
\frac{1-r+\sqrt{1+r^2}}{1+r-\sqrt{1+r^2}}=\frac{2m_2+1}{2m_1+1} ,  \\
\frac{-1+r+\sqrt{1+r^2}}{1+r-\sqrt{1+r^2}}=\frac{2m_3+1}{2m_1+1} ,
\label{four_ferm_ratio}
\end{eqnarray*}
which determine the ratio $r=J_2/J_1$ for perfect state transfer.
These equations can only be satisfied if 
\begin{eqnarray}
2m_1 &=& -2-m_2-m_3 \nonumber \\ 
& & +\sqrt{2+4m_2+m_2^2+4m_3+6m_2m_3+m_3^2} \nonumber \\ 
\mathrm{or} & & \label{condition} \\
2m_1 &=& m_2+m_3 \nonumber \\ 
& & + \sqrt{2+4m_2+m_2^2+4m_3+6m_2m_3+m_3^2}.\nonumber 
\end{eqnarray}
Since $m_1$ is integer, $\sqrt{2+4m_2+m_2^2+4m_3+6m_2m_3+m_3^2}=k$
should also be some integer $k$. First notice that 
\begin{eqnarray*}
2+4m_2+m_2^2+4m_3+6m_2m_3+m_3^2 \\
= 2(m_2+m_3+1)^2-(m_2-m_3)^2.
\end{eqnarray*}
We now prove that the condition
\begin{equation}
k^2+(m_2-m_3)^2 = 2(m_2+m_3+1)^2 \label{eq-cond}
\end{equation}
cannot be satisfied with any set of integers $m_2,m_3$ and $k$.
There are four possible cases: 
(i) $k$ is odd and $(m_2-m_3)$ is even, 
(ii) $k$ is even and $(m_2-m_3)$ is odd,
(iii) both $k$ and $(m_2-m_3)$ are odd, and 
(iv) both $k$ and $(m_2 - m_3)$ are even.
Note that if $(m_2 - m_3)$ is odd (even) then $(m_2+m_3 +1)$ is even (odd).
Cases (i) and (ii) are then ruled out since they yield odd left-hand side (lhs) 
of Eq.~\eqref{eq-cond}, whereas the right-hand side (rhs) is always even. 
For case (iii) the rhs is divisible by $4$ without remainder and the lhs is not. 
Finally, for case (iv) the lhs is divisible by $4$ without remainder and the
rhs is not. This means that conditions \eqref{condition} cannot 
be satisfied. Hence, an isotropic ($XXX$) spin-chain cannot realize 
perfect state transfer, unless the diagonal elements of the Hamiltonian 
matrix in Eq.~\eqref{SupEq:HamSpinMartix} are modified by a local 
(magnetic field) perturbation, cf. Eq.~\eqref{Eq:HamSpinFerm} below.

\subsection{Bosons} 
\label{subsec:appABos}

We now consider bosons with $\kappa=1$ leading to the following eigenvalues 
and eigenvectors 
\begin{widetext}
\begin{eqnarray*} 
\lambda^{(\mathrm{b})}_1 = 3J_1+J_2-\sqrt{J_1^2+J_2^2}, & \quad & 
\ket{\Psi_1^{(\mathrm{b})}} = \frac{1}{2 \sqrt{J_1^2+J_2^2 + J_2 \sqrt{J_1^2+J_2^2} }}\{J_1, J_2+\sqrt{J_1^2+J_2^2},J_2+\sqrt{J_1^2+J_2^2},J_1 \} ; \nonumber \\
\lambda^{(\mathrm{b})}_2 = 2 J_1+ 2J_2, & & \ket{\Psi_2^{(\mathrm{b})}} = \frac{1}{2}\{-1,-1,1,1\} ; \nonumber \\
\lambda^{(\mathrm{b})}_3 = 3J_1+J_2+\sqrt{J_1^2+J_2^2}), &  & \ket{\Psi_3^{(\mathrm{b})}}=\frac{1}{2 \sqrt{J_1^2+J_2^2 - J_2 \sqrt{J_1^2+J_2^2} }}\{J_1,J_2-\sqrt{J_1^2+J_2^2},J_2-\sqrt{J_1^2+J_2^2},J_1\} ; \nonumber \\
\lambda^{(\mathrm{b})}_4 = 4J_1+2J_2, & & \ket{\Psi_4^{(\mathrm{b})}} = \frac{1}{2}\{-1,1,-1,1\}.
\end{eqnarray*}
The initial state $\ket{\Psi^{(\mathrm{b})}_{\mathrm{in}}} = \ket{\!\ds \us \us \us}$
now evolves as
\begin{eqnarray*}
\ket{\Psi^{(\mathrm{b})}(t)} &=& \frac{J_1}{2 \sqrt{J_1^2+J_2^2 + J_2 \sqrt{J_1^2+J_2^2} }} \ket{\Psi_1^{(\mathrm{b})}} e^{-i \lambda^{(\mathrm{b})}_1 t}
- \frac{1}{2} \ket{\Psi_2^{(\mathrm{b})}} e^{-i \lambda^{(\mathrm{b})}_2 t} \\ & &
+ \frac{J_1}{2 \sqrt{J_1^2+J_2^2 - J_2 \sqrt{J_1^2+J_2^2} }} \ket{\Psi_3^{(\mathrm{b})}} e^{-i \lambda^{(\mathrm{b})}_3 t} 
- \frac{1}{2} \ket{\Psi_4^{(\mathrm{b})}} e^{-i \lambda^{(\mathrm{b})}_4 t} .
\end{eqnarray*}
\end{widetext}
Note that 
$\lambda^{(\mathrm{b})}_4-\lambda^{(\mathrm{b})}_3=\lambda^{(\mathrm{f})}_2$,
$\lambda^{(\mathrm{b})}_4-\lambda^{(\mathrm{b})}_2=\lambda^{(\mathrm{f})}_3$, and 
$\lambda^{(\mathrm{b})}_4-\lambda^{(\mathrm{b})}_1=\lambda^{(\mathrm{f})}_4$.
As a result, the fidelity of state transfer, 
$F(t) \equiv |\braket{\Psi(t)}{\!\us \us \us \ds}|^2$,
is the same for both fermions ($\kappa \to \infty$) and bosons with $\kappa=1$,
which holds true for $N_\downarrow=1$ and any $N_\uparrow$.

Next, in the special case of perfect state transfer, $J_2/J_1 = \sqrt{4/3}$, 
with the Heisenberg $XX$ model, $\kappa=2$, we have the equidistant spectrum 
$\lambda^{(\mathrm{b})}_1= \frac{\sqrt{12}-1}{2} J_2$, 
$\lambda^{(\mathrm{b})}_2 = \lambda^{(\mathrm{b})}_1 + J_2$, 
$\lambda^{(\mathrm{b})}_3 = \lambda^{(\mathrm{b})}_1 + 2J_2$, 
and $\lambda^{(\mathrm{b})}_4=\lambda^{(\mathrm{b})}_1+3J_2$, 
leading to the fastest transfer time $t_{\mathrm{out}} = \pi/J_2$.

The final case discussed in the text concerns the limit $\kappa \ll 1$ when 
inter-species interaction is much larger than the intra-species interaction. 
Then the two lowest eigenstates 
$\ket{\Psi^{(\mathrm{b})}_1} \simeq \frac{1}{\sqrt{2}}\{1,0,0,1\}$
and $\ket{\Psi^{(\mathrm{b})}_2} \simeq \frac{1}{\sqrt{2}}\{-1,0,0,1\}$ become 
degenerate, $\lambda^{(\mathrm{b})}_1 \simeq \lambda^{(\mathrm{b})}_2$ and separated 
from the other two eigenstates
$\ket{\Psi^{(\mathrm{b})}_3} \simeq \frac{1}{\sqrt{2}}\{0,1,1,0\}$ and
$\ket{\Psi^{(\mathrm{b})}_4} \simeq \frac{1}{\sqrt{2}}\{0,-1,1,0\}$ by
$\lambda^{(\mathrm{b})}_{3,4} - \lambda^{(\mathrm{b})}_{1} \simeq \frac{2J_2}{\kappa} \pm J_2$. The state transfer between the initial $\ket{\!\ds \us \us \us}$ and 
final $\ket{\!\us \us \us \ds}$ states proceeds then via non-resonant 
intermediate states $\ket{\!\us \ds \us \us}$ and $\ket{\!\us \us \ds \us}$ 
as a third-order process with the effective Rabi frequency 
$J_{\mathrm{eff}} \simeq \frac{J_1 J_2 J_1}{(2J_2/\kappa)^2} 
= \frac{\kappa^2 J_1^2}{4 J_2}$.

\section{Effective spin model in a magnetic field}
\label{sec:appB}

Here we outline the derivation of the effective spin Hamiltonian $\tilde{H}_s$
for $N$ particles in an (effective) external magnetic field $B(x)\sigma_z$. 
We consider a single spin-down particle $N_{\ds} = 1$ and assume 
weak magnetic field $B(x)=b(x)/g$ ($g \gg 1$) which modifies 
the Hamiltonian $H$ of Eq.~(1) as  
\begin{equation}
\tilde{H} = H + \sum_{i=1}^{N-1} \frac{b(x_{\us,i})}{g} - \frac{b(x_{\ds,1})}{g} .
\end{equation}
For the corresponding energy of $N$-particle eigenfunction $\Psi$, 
to linear order in $1/g$, we then obtain
\begin{equation}
\tilde{E} = E -2 \frac{\sum_{j=1}^{N} \frac{\beta_j}{g} a_{j}^2}
{\sum_{j=1}^{N} a_j^2} + \sum_{j=1}^{N} \beta_j ,
\end{equation} 
where we write simply $a_j$ instead of $a_{j|k}$ for a single impurity 
(spin-down) particle, while the geometric factors are
\begin{equation}
\beta_j = \frac{\int \prod_{i=1}^{N-1} d x_{\us,i} \, d x_{\ds,1} 
|\Phi_0|^2 \, b(x_{\ds,1})} 
{\int \prod_{i=1}^{N-1} d x_{\us,i} \, d x_{\ds,1}  
|\Phi_0(\{ : x_{\us,i} , x_{\ds,1} : \})|^2 } ,
\end{equation}
where in $\Phi_0$ in the numerator the spin-down particle $x_{\ds,1}$ 
is placed at position $j$. 
The effective spin Hamiltonian for the case $\kappa \to \infty$ 
can now be cast as
\begin{equation}
\tilde{H}_s = E_0 \mathbf{I} - \hlf \sum_{j=1}^{N-1} 
J_j (\bs{\sigma}^j \bs{\sigma}^{j+1} - \mathbf{I} ) + \sum_{j=1}^{N} h_j \sigma_z^j,
\label{Eq:HamSpinFerm}
\end{equation}
with $h_j = \beta_j$.

\end{document}